\documentclass[prb,twocolumn,showpacs]{revtex4}
\usepackage{color,soul}
\usepackage{amsmath}
\usepackage{dcolumn}
\usepackage{graphicx}
\usepackage{bm}
\usepackage{amssymb}
\usepackage{subfigure}
\begin{document}

\title{Can Hubbard model resist electric current?}
\author{Tao Li and Jianhua Yang}
\affiliation{Department of Physics, Renmin University of China, Beijing 100872, P.R.China}

\begin{abstract}
It is claimed in a recent quantum Monte Carlo simulation study that the linear-in-T dc resistivity observed in the strange metal phase of the high-T$_{c}$ cuprate superconductors can be reproduced in the pure two dimensional Hubbard model\cite{Huang}. Here we show that such a translational invariant electronic model can not support a steady state current in the presence of a uniform electric field at any finite temperature and thus does not posses a well defined dc resistivity at a general incommensurate electron filling. Extrinsic scattering processes must be invoked to bypass the momentum and the energy bottleneck to establish such a steady state current. We argue that a holographic transport picture, in which the Drude weight acts as an emergent low energy degree of freedom, may be more appropriate for the understanding of the strange metal behavior of the cuprate superconductors. Within such a transport picture, the Hubbard interaction is mainly responsible for the reduction of the Drude weight participating the dc transport but is insufficient to resolve the momentum and energy bottleneck in the dc transport process alone. It is the interplay between the Umklapp scattering caused by the Hubbard interaction and other scattering processes like the electron-phonon scattering and impurity scattering that determine the transport relaxation rate.
\end{abstract}

\maketitle

The linear-in-temperature dc resistivity in the so called strange metal phase is arguably the most prominent manifestation of the non-Fermi liquid behavior of the high-T$_{c}$ cuprate superconductors\cite{Cooper,Taillefer,Proust,Hussey,Phillips}. Different from conventional Fermi liquid system such as the elemental metals, in which the dc resistivity can be attributed to the electron-phonon scattering and thus exhibits clear $T/\omega_{D}$ scaling, with $\omega_{D}$ the Debye frequency of the system, the dc resistivity of the high-T$_{c}$ cuprate superconductors in the strange metal phase is a perfect linear function of temperature from very high temperature all the way down to zero temperature, without any sign of an additional characteristic temperature scale. Such a strange metal behavior is anomalous from the Fermi liquid theory perspective in the following three aspects. 

The first anomaly concerns the behavior of $\rho(T)$ in the low temperature limit. It is found that the linear temperature dependence of $\rho(T)$ extends directly to the lowest temperature measured. On the other hand, according to the Fermi liquid theory we expect either a $T^{5}$ dependence in $\rho(T)$ in the low temperature limit if it is dominated by electron-phonon scattering, or a $T^{2}$ dependence if it is dominated by electron-electron scattering. Note that the scattering caused by impurities only contributes a constant shift to the resistivity if the quasiparticle transport picture apply. Some people proposed that such an anomalous low-T behavior in $\rho(T)$ may be attributed to the scattering from some exotic gapless collective fluctuation of the electron system, in particular the critical fluctuation around a quantum critical point. However, the linear-in-T resistivity in the low temperature limit is found to exist in broad doping range of phase diagram of the cuprate superconductors\cite{Cooper,Taillefer,Proust,Hussey,Phillips}.

The second anomaly concerns the behavior of $\rho(T)$ in the intermediate temperature range. According to the Fermi liquid theory perspective, the quasiparticle scattering rate contributed by different scattering channels should be added up to determine the dc resistivity. Thus, if the low temperature resistivity is dominated by some exotic scattering mechanism different from the electron-phonon coupling, we should expect a change in the temperature dependence of the dc resistivity at the temperature when the electron-phonon scattering starts to play an essential role. Such an expected turning point is however never observed in the resistivity curve of the cuprate superconductors. The rule of addition of quasiparticle scattering rate in the Fermi liquid theory, namely the Matthiessen's rule, seems to be inapplicable here.

The third anomaly concerns the behavior of $\rho(T)$ in the high temperature limit. According to the Fermi liquid theory, one should expect saturation of the resistivity when the quasiparticle mean free path becomes comparable to the lattice constant. However, the cuprate superconductors does not exhibit any hint of change in the slope of the resistivity at temperature as high as 1000K. The existence of the so called bad metal regime calls for a serious rethinking of the quasiparticle transport picture of the Fermi liquid theory.
    
It is generally believed that the electron correlation effect plays an essential role in the origin of the strange metal behavior. However, it is not clear if such an anomalous transport behavior can appear in a purely electronic model with translational symmetry. The answer to such a question is important for several reasons. In fact, it has been argued that nontrivial conclusion can be drawn if the strange metal behavior appear as an intrinsic property of a translational invariant electron model at a general incommensurate electron filling. Based on the analysis of the emergent symmetry in the low energy effective theory and its related anomaly, it is argued in Ref.[\onlinecite{Else1,Else2,Else3}] that the linear-in-T resistivity can emerge in a translational invariant system in the low temperature limit only when the susceptibility of a quantity with the same symmetry character as the loop current order proposed by Varma\cite{loop} diverges. Such a peculiar mechanism for the generation of dc resistivity is called the critical drag mechanism\cite{Else3}. Leaving aside the validity of such an argument, the answer to this question is also of fundamental importance when we attempt to construct a transport theory beyond the quasiparticle picture, more specifically, to understand how dc current dissipate without resorting to the quasiparticle picture.

Recently,  a quantum Monte Carlo simulation of the two dimensional Hubbard model on the square lattice indicates that such a translational invariant electron model may indeed posses a linear-in-T resistivity at a general electron filling\cite{Huang}. In this simulation, the authors have calculated the regular part of the optical conductivity $\sigma^{reg}(\omega)$ of the Hubbard model. The zero frequency limit of $\sigma^{reg}(\omega)$ is then taken as the dc conductivity of the system. While it is impossible to reach temperature significantly lower than the hopping integral of the Hubbard model as a result of the severe sign problem in the quantum Monte Carlo simulation, their numerical results do suggest the persistence of the linear-in-T behavior at lower temperature. Such a result seems to be at odd with the argument presented in Ref.[\onlinecite{Else1,Else2,Else3}], since we do not expect loop current ordering instability to occur in the Hubbard model at a general electron filling and with a general $U$. 

The purpose of this paper is to question the result of such a numerical simulation. We argue that no meaningful dc resistivity can be defined at any finite temperature for the translational invariant Hubbard model at a general incommensurate electron filling. In fact, a well defined dc resistivity can be defined only when a steady state current can be established in the presence of a uniform static electric field. For the translational invariant Hubbard model, we argue that this is impossible as a result of insufficient momentum relaxation and energy relaxation. We arrive at our conclusion through two routes. We first present an analysis based on the linear response theory. This is then followed by an analysis on the bottleneck effect to establish a steady state current in the presence of a uniform static electric field, which is the prerequisite to define a meaningful resistivity. 

The Hubbard model we study is given by
\begin{equation}
H=-t\sum_{<i,j>,\sigma}(c^{\dagger}_{i,\sigma}c_{j,\sigma}+h.c.)+U\sum_{i}n_{i,\uparrow}n_{i,\downarrow}-\mu\sum_{i,\sigma}n_{i,\sigma},
\end{equation}
and is defined on the square lattice with periodic boundary condition in both the $x$- and the $y$-direction. $t$ denotes the hopping integral between nearest neighboring sites. The uniform electric field is generated by a linearly time dependent vector potential $\mathbf{A}(t)$ along the $x$-direction of the two dimensional lattice. The two components of $\mathbf{A}(t)$ are given by
\begin{equation}
A_{x}(t)=-Et;\ A_{y}(t)=0.
\end{equation} 
$\mathbf{A}$ is coupled to the electron hopping term through the Peierls substitution of the form
\begin{eqnarray}
H=-&t&\sum_{<i,j>,\sigma}(e^{-i \mathbf{A}(t) \cdot ({\mathbf{r}_{i}-\mathbf{r}_{j}})} c^{\dagger}_{i,\sigma}c_{j,\sigma}+h.c.)\nonumber\\
+&U&\sum_{i}n_{i,\uparrow}n_{i,\downarrow}-\mu\sum_{i,\sigma}n_{i,\sigma}
\end{eqnarray}
Here we have used the natural unit in which $e=c=\hbar=1$ for convenience.

Coupling to the electric field through the Peierls substitution has the advantage of preserving the translational symmetry of the original electronic model. After the Fourier transformation we can rewrite the model as follows
\begin{equation}
H=H_{K}+H_{U},
\end{equation}
in which 
\begin{equation}
H_{K}=\sum_{\mathbf{k},\sigma}\epsilon_{\mathbf{k}-\mathbf{A}(t)}c^{\dagger}_{\mathbf{k},\sigma}c_{\mathbf{k},\sigma}.
\end{equation}
Here $\epsilon_{\mathbf{k}}=-2t(\cos k_{x}+\cos k_{y})-\mu$ is the single particle dispersion on the square lattice.
\begin{equation}
H_{U}=U\sum_{i}n_{i,\uparrow}n_{i,\downarrow}
\end{equation} 
in the Hubbard interaction.

In the linear response regime, the electric conductivity of the model is given by the Kubo formula as follows
\begin{equation}
\sigma^{xx}(\omega+i0^{+})=\frac{i}{\omega+i0^{+}}[\ |\langle K_{x} \rangle|+\Pi_{xx}(\omega+i0^{+})\ ].
\end{equation}
Here $K_{x}=-t\sum_{i,\sigma}(c^{\dagger}_{i+\delta_{x},\sigma}c_{i,\sigma}+h.c.)$ is the electron kinetic energy operator in the $x$-direction. $\Pi_{xx}(\omega+i0^{+})$ is the retarded current-current correlation function defined as
\begin{equation}
\Pi_{xx}(\omega)=-i\int_{0}^{\infty}dt\ e^{i\omega t}\langle [j_{x}(t),j_{x}(0)]\rangle,
\end{equation}
in which $j_{x}=-it\sum_{i,\sigma}(c^{\dagger}_{i+\delta_{x},\sigma}c_{i,\sigma}-h.c.)$ is the $x$-component of the electric current for $\mathbf{A}=0$, which has the form of
\begin{equation}
j_{x}=\sum_{\mathbf{k},\sigma}v^{x}_{\mathbf{k}}c^{\dagger}_{\mathbf{k},\sigma}c_{\mathbf{k},\sigma}
\end{equation}
in the Fourier space. Here $v^{x}_{\mathbf{k}}=\frac{\partial \epsilon_{\mathbf{k}}}{\partial k_{x}}$ is the group velocity of the band electron. For finite $\mathbf{A}$ the electric current operator becomes
\begin{equation}
j_{x}(\mathbf{A})=\sum_{\mathbf{k},\sigma}v^{x}_{\mathbf{k}-\mathbf{A}}c^{\dagger}_{\mathbf{k},\sigma}c_{\mathbf{k},\sigma}.
\end{equation}
 Using the relation $\frac{1}{x+i0^{+}}=\frac{1}{x}-i\delta(x)$ we have
\begin{equation}
\mathrm{Re}\sigma^{xx}(\omega)=D\delta(\omega)+\sigma^{reg}(\omega),
\end{equation}
in which
\begin{equation}
D=|\langle K_{x} \rangle|+\mathrm{Re}\Pi_{xx}(i0^{+})
\end{equation}
is the Drude weight of the electric conductivity.
\begin{equation}
\sigma^{reg}(\omega)=-\frac{\mathrm{Im}\Pi_{xx}(\omega+i0^{+})}{\omega}
\end{equation}
is the optical absorption caused by electron transition during the current coupling.
Using the Kronig-Kramers relation satisfied by the retarded current-current correlation function, we have
\begin{eqnarray}
\mathrm{Re}\Pi_{xx}(i0^{+})=\frac{2}{\pi}\int_{0^{+}}^{\infty}d\omega \frac{\mathrm{Im}\Pi_{xx}(\omega+i0^{+})}{\omega}\nonumber\\
=-\frac{2}{\pi}\int_{0^{+}}^{\infty}d\omega\sigma^{reg}(\omega).
\end{eqnarray}
Thus we have 
\begin{equation}
D=|\langle K_{x} \rangle|-\frac{2}{\pi}\int_{0^{+}}^{\infty}d\omega\sigma^{reg}(\omega).
\end{equation}
The physical meaning of this equation is self-evident: the total spectral weight involved in the dissipationless dc transport of the electron system in the $x$-direction is given by the total kinetic energy of the electron system in the $x$-direction subtracting the spectral weight involved in the optical absorption at nonzero frequency, which is a manifestation of the electron incoherence induced by various scattering processes.    

In the absence of the Hubbard interaction, $j_{x}$ is a conserved quantity. We thus have $\mathrm{Im}\Pi_{xx}(\omega+i0^{+})=0$ and $D=|\langle K_{x} \rangle|$. The Hubbard model is thus perfectly conducting in the linear response regime at any finite temperature when $U=0$. In fact, it can be shown that the system can not establish a steady state current, but would rather execute Bloch oscillation in the presence of a finite uniform electric field. This can be seen simply from the fact that the lattice momentum of electron will drift with time in the electric field as 
\begin{equation}
\frac{d\mathbf{k}}{dt}=\mathbf{E}
\end{equation} 
and that there is no other momentum relaxation channel when $U=0$. In a more formal way, this can be seen from the time evolution of the density matrix of the system, which is governed by the Liouville equation of the form
\begin{equation}
i\frac{\partial \hat{\rho}(t)}{\partial t}=[\hat{\rho}(t), H(t)].
\end{equation}  
Now we assume that at $t=0$ the system is in the thermal equilibrium, namely
\begin{equation}
\hat{\rho}(t=0)=\frac{1}{Z}e^{-\beta \sum_{\mathbf{k},\sigma}\epsilon_{\mathbf{k}}c^{\dagger}_{\mathbf{k},\sigma}c_{\mathbf{k},\sigma}},
\end{equation}
in which $Z$ denotes the partition function of the system and $\beta=1/k_{B}T$. Since $[\hat{\rho}(t), H(t)]=0$ when $U=0$, we have 
\begin{equation}
\hat{\rho}(t)\equiv\hat{\rho}(t=0).
\end{equation}
The current induced by the electric field is thus given by
\begin{eqnarray}
\langle j_{x}(\mathbf{A}) \rangle&=& \mathrm{Tr} [ \hat{\rho}(t)  j_{x}(\mathbf{A}) ]= \mathrm{Tr} [ \hat{\rho}(t=0)  j_{x}(\mathbf{A}) ]\nonumber\\
&=&2\sum_{\mathbf{k}}v^{x}_{\mathbf{k}-\mathbf{A}}n_{F}(\epsilon_{\mathbf{k}})\nonumber\\
&=&\langle K_{x} \rangle \sin A_{x}(t)\nonumber\\
&=&-\langle K_{x} \rangle \sin Et.\nonumber\\
&=&D \sin Et.
\end{eqnarray}
and thus will execute Bloch oscillation.

Now we turn on the Hubbard interaction term $H_{U}$. We assume that the perturbation expansion in $U$ converges at a general incommensurate band filling. Let us estimate how the Drude weight $D$ evolve with $U$. To the lowest order in the perturbative expansion in $U$, it is obvious that the expectation value of the kinetic energy $\langle K_{x}\rangle$ should depend linearly on $U$, namely
\begin{equation}
|\langle K_{x} \rangle|=|\langle K_{x} \rangle|_{U=0}-\alpha U
\end{equation}
with $\alpha$ some positive constant. An estimation of $\mathrm{Re}\Pi_{xx}(i0^{+})$, namely, the suppression of the Drude weight due to the dissipative optical absorption is more difficult.  However, in the high temperature limit we have
\begin{eqnarray}
\mathrm{Re}\Pi_{xx}(i0^{+})&=&\frac{2}{\pi}\int_{0^{+}}^{\infty}d\omega \frac{\mathrm{Im}\Pi_{xx}(\omega+i0^{+})}{\omega}\nonumber\\
&=&-\frac{1}{\pi}\int_{0^{+}}^{\infty}d\omega \frac{R(\omega)}{\omega}\nonumber\\
&=&-\frac{1}{\pi}\int_{0^{+}}^{\infty}d\omega (1-e^{-\beta\omega})\frac{J(\omega)}{\omega}\nonumber\\
&\approx&-\frac{\beta}{\pi}\int_{0^{+}}^{\infty}d\omega J(\omega)\nonumber\\
&=&-\frac{\beta}{\pi}\langle j^{2}_{x} \rangle
\end{eqnarray}
in which $R(\omega)=-2\mathrm{Im}\Pi_{xx}(\omega+i0^{+})$ denotes the spectral function of  the current operator. $J(\omega)$ is the dynamical structure factor of the current operator and is given by
\begin{equation}
J(\omega)=\int_{-\infty}^{\infty} dt \langle j_{x}(t) j_{x}(0)\rangle e^{i\omega t}
\end{equation}
It is clear that $\langle j^{2}_{x} \rangle$ should depend linearly on $U$ for small $U$. We thus expect $\mathrm{Re}\Pi_{xx}(i0^{+})$ to increase linearly with $U$ for small $U$ at sufficiently high temperature. Thus the Drude weight should decrease linearly with $U$ for small $U$ at sufficiently high temperature. 

On the other hand, at zero temperature we have 
\begin{eqnarray}
\mathrm{Re}\Pi_{xx}(i0^{+})&=&\frac{2}{\pi}\int_{0^{+}}^{\infty}d\omega \frac{\mathrm{Im}\Pi_{xx}(\omega+i0^{+})}{\omega}\nonumber\\
&=&-\frac{1}{\pi}\int_{0^{+}}^{\infty}d\omega \frac{J(\omega)}{\omega}\nonumber\\
&=&-\frac{\langle \omega^{-1} \rangle_{J}}{\pi}\langle j^{2}_{x} \rangle 
\end{eqnarray}
in which 
\begin{equation}
\langle \omega^{-1} \rangle_{J}=\frac{\int_{0^{+}}^{\infty}d\omega \ \omega^{-1} \ J(\omega)}{\int_{0^{+}}^{\infty}d\omega J(\omega)}
\end{equation}
is the weighted average of $\omega^{-1}$ over the distribution $J(\omega)$. If we assume that the functional form of $J(\omega)$ does not change significantly with $U$, then 
$\mathrm{Re}\Pi_{xx}(i0^{+})$ should still increase linearly with $U$. 

Thus we expect that the Drude weight to decrease linearly with $U$ for small $U$ at a general incommensurate filling at any temperature, or
\begin{equation}
D(U,T)=D(U=0,T)-\gamma(T) U+...
\end{equation}
in which $\gamma(T)$ is temperature dependent constant. The Drude weight is thus nonzero at any temperature in the small $U$ regime. 

We note that the existence of a nonzero Drude weight has also been argued from the perspective of emergent symmetry in the low energy effective theory and related anomaly matching at a general incommensurate filling\cite{Else1}. Such an argument is thought to be valid even when perturbation expansion about $U$ fails. 
However, we note that the argument presented in Ref.[\onlinecite{Else1}] is based on emergent symmetry in the low energy effective theory, which is valid in the strict sense only in the zero temperature limit. Irrelevant couplings at finite energy or finite temperature may still render the Drude weight a finite life time. What we have argued here is that for a purely electronic model with translational symmetry, while the magnitude of the Drude weight can be reduced by electron correlation effect, it remains dissipationless even at finite temperature.

In the presence of a uniform and static electric field, the Drude weight would execute Bloch oscillation in the background of periodic lattice as we have shown for the noninteracting case. We thus do not expect the Hubbard model to establish a steady state current in the presence of a finite uniform electric field at small $U$. This can be seen more generally from the analysis of the bottleneck effect to establish such a steady state current. In the presence of the uniform electric field, the lattice momentum of the electron will drift with a constant velocity in the Brillouin zone, namely
\begin{equation}
\frac{d\mathbf{k}}{dt}=\mathbf{E}
\end{equation} 
We note that this is an exact equation originated from the requirement of gauge invariance and does not depend on any microscopic detail of the model. We also note that in the low energy effective theory description such a uniform drift corresponds just to the anomaly related to the emergent symmetry on the Fermi surface. On the other hand, the relaxation rate of lattice momentum caused by the Umklapp scattering process should be proportional to $U$ at small $U$. A steady state current can be established only when the increase of the lattice momentum in the electric field can be balanced by the Umklapp scattering process. This is clearly impossible for general value of $U$ and $E$. This is the momentum bottleneck to establish a steady state current for the Hubbard model in a uniform electric field. Moreover, a steady state current in the direction of the applied dc electric field would imply a constant rate of the generation of the Joule heat in the system, namely
\begin{equation}
\frac{dw}{dt}=\mathbf{J}\cdot \mathbf{E}
\end{equation}
Such Joule heat must be efficiently dissipated to establish the steady state current. This is clearly impossible in a model involving only electron degree of freedom. Since a steady state current is the prerequisite to define dc resistivity, we think no meaningful dc resistivity can be defined for the Hubbard model at any temperature as a result of the momentum and energy bottleneck effect mentioned above. Since our argument relies only on the translational invariant nature of the model and the gauge principle, our conclusion should apply for more general pure electronic models, such as the $t-J$ model. It is interesting and important to check these postulates with numerical simulation on the Hubbard and the $t-J$ model.

The discussion above suggests that it is better to interpret the Drude weight $D$ as the total spectral weight participating the dc transport, rather than a measure of the strength of the dissipationless current response. With such an understanding in mind, the problem is then how the Drude weight dissipate its momentum and energy to establish a steady state current in the presence of a uniform dc electric field. In the Landau Fermi liquid theory perspective, the dissipation of the dc transport current is attributed to the dissipation of the individual quasiparticles carrying the current. The dc resistivity is thus determined by the total scattering rate of the quasiparticles, which is the sum of the scattering rate from different scattering channels, in accordance with the Matthiessen's rule. The analysis presented in this paper shows that while the magnitude of the Drude weight can be suppressed by the interaction term in the Hubbard model, the Umklapp scattering process alone is in general insufficient to dissipate the lattice momentum accumulated in a uniform electric field. Extrinsic scattering process, such as the electron-phonon scattering or the impurity scattering, must be invoked to bypass both the momentum and the energy bottleneck to establish a steady state current for the Drude weight.

In the quantum Monte Carlo simulation study of the Hubbard model, the zero frequency limit of $\sigma^{reg}(\omega)$ is taken as dc conductivity of the system. This only make sense if the Drude weight is identically zero. As we argued above, the Drude weight is in general nonzero at a general incommensurate electron filling for a translational invariant electronic model. In fact, the nonzero Drude weight is a direct manifestation of the anomaly related to the emergent symmetry in the low energy effective theory. Thus, taking the zero frequency limit of $\sigma^{reg}(\omega)$ is not faithful representation of the dc transport of a translational invariant electronic model. Nevertheless, the computation of $\sigma^{reg}(\omega)$ is helpful for us to deduce the reduction of the Drude weight from the dissipative optical absorption.  

As implied by the three anomalies about the strange metal behavior summarized at the beginning of this paper, the conventional transport theory based on the quasiparticle picture should be abandoned in the cuprate superconductors. A thorough understanding of the relaxation mechanism of the dc current without resorting to the quasiparticle picture is a challenging theoretical task. Nevertheless, even in the absence of well defined quasiparticle the relaxation rate of the dc current is still well defined. It manifests itself simply in the broadening of the Drude peak in the optical conductivity spectrum. Indeed, recent optical measurement indicates that the optical spectrum of the cuprate superconductors can be thought of as the sum of a well defined Drude peak at low frequency and a conformal tail at higher frequency\cite{Heuman}. Thus, it is convenient to take the Drude weight as an emergent low energy degree of freedom when we go beyond the quasiparticle transport picture\cite{Hartnoll}. Within such a holographic transport picture, the Umkalpp scattering process is mainly responsible for the reduction of the Drude weight but is in general insufficient to resolve the momentum and energy bottleneck in the dc transport process alone. The interplay between the Umklapp scattering process and other extrinsic scattering processes in determining the relaxation rate of the Drude weight can be very complicated and interesting. One thing that one can naturally expect within such a holographic transport picture is that the Matthiessen's rule may not be valid any more. 

In summary, we show that it is in general impossible to define a meaningful dc resistivity for the translational invariant Hubbard model at any temperature as a result of insufficient momentum and energy relaxation of the Drude weight in the presence of a uniform electric field. We argue that a holographic transport picture, in which the Drude weight acts as an emergent low energy degree of freedom, may be more appropriate for the understanding of the strange metal behavior of the cuprate superconductors. Within such a novel transport picture, the Hubbard interaction is mainly responsible for the reduction of the Drude weight participating the dc transport. It is the interplay between the Umklapp scattering caused by the Hubbard interaction and other scattering processes like the electron-phonon scattering and impurity scattering that determine the transport relaxation rate.

 \begin{acknowledgments}
We acknowledge the support from the grant NSFC 12274453. Tao Li would like to acknowledge the invaluable support from Prof. Chun-Fang Li.
\end{acknowledgments}

\end{document}